\documentclass[10pt,a4paper]{article}

\usepackage[utf8]{inputenc}
\usepackage{amsmath}
\usepackage{amsfonts}
\usepackage{amssymb}
\usepackage[dvips,letterpaper,margin=0.75in,bottom=0.5in]{geometry}

\usepackage[T1]{fontenc}
\usepackage[utf8]{inputenc}
\usepackage{authblk}
\usepackage{scrextend}
\usepackage{footmisc}
\usepackage{indentfirst}

\title{Sections and Chapters}
\usepackage{cite}
\newcommand{\ar}{u_{+}}
\newcommand{\al}{u_{-}}
\newcommand{\ara}{u^{\dagger}}
\newcommand{\ala}{u}

\newcommand{\Pm}[2]{u^{\dagger\/{#1}}u^{#2}}
\newcommand{\Pmt}[2]{u^{#1}u^{\dagger\/{#2}}}
\newcommand{\Pmm}{u^{\dagger\/m}u^{m}}
\newcommand{\Pmb}{u^{\dagger}u}
\newcommand{\Pmn}{u^{\dagger\/n}u^{n}}

\newcommand{\agb}{u_{-}u_{+}=1-u_{+}^{d-1}u_{-}^{d-1}}
\newcommand{\agi}{u_{+}u_{-}=1-u_{-}^{d-1}u_{+}^{d-1}}

\newcommand{\abs}[1]{\left| #1 \right|} 
 
\newcommand{\ket}[1]{\left| #1 \right>} 
\newcommand{\braket}[2]{\left< #1 \vphantom{#2} \right|
 \left. #2 \vphantom{#1} \right>} 
\let\baraccent=\= 
\renewcommand{\=}[1]{\stackrel{#1}{=}} 

\title{Quantum Mechanics in a Space with a Finite Number of Points }
\author{Metin Arik$^{1,}$\footnote{e-mail:metin.arik@boun.edu.tr}}
\author{Medine Ildes$^{1,}$\footnote{e-mail:medine.ildes@boun.edu.tr}}
\affil{$^{1}$Department of Physics, Bogazici University, Bebek, Istanbul, Turkiye}

\begin{document}

\maketitle

\begin{abstract}
We define a deformed kinetic energy operator for a discrete position space with a finite number of points. The structure may be either periodic or nonperiodic with well-defined end points. It is shown that for the nonperiodic case the translation operator becomes nonunitary due to the end points. This uniquely defines an algebra which has the desired unique representation. Energy eigenvalues and energy wave functions for both cases are found. As expected, in the continuum limit the solution for the nonperiodic case becomes the same as the solution of an infinite one dimensional square well and the periodic case solution becomes the same as the solution of a particle in a box with periodic boundary conditions.
\end{abstract}

\section{Introduction}
\indent The conventional formulation \cite{john1955} of quantum mechanics starts with a position space which is the set of real numbers. This leads to physical states which are vectors of a separable Hilbert space. The observables are then described by hermitian operators on this Hilbert space. Angular momentum and spin one half are the well known systems in which observables have finite and discrete values. Furthermore, quantum information theory, quantum optics and the lattice models are all realized in a finite dimensional Hilbert space. One can find a guide to the literature on the applications in Vourdas' work \cite{0034-4885-67-3-R03}, where quantum systems with finite Hilbert space are considered and phase-space methods are discussed. In this paper we would like to investigate what happens when the position space or the momentum space consists of a finite number of points so that the Hilbert space associated with the quantum mechanics is finite dimensional. Schwinger \cite{schwinger1960unitary} has given a most instructive example where the position space and the momentum space each consist of $d$ points and are both periodic. The starting point for this consideration is the well-known simple quantum mechanics on the circle $ S^{1} $. In this case the position space is continuous whereas the momenta are given by $ p_{n}= \dfrac{\hbar}{r} n, (n=0,\pm1,\pm2,...) $ where r is the radius of the circle. Now restricting the position space $S^{1}$ to integer multiples of angle $ 2\pi/d $, the position eigenvectors can be denoted by $  \ket{n}, n=0,1,2,...,d-1$  such that
\begin{align}  
X\ket{n} =\frac{ 2\pi\/r}{d}\/n\ket{n} .
\end{align}
\\Note that this equation is not well-defined because the operator X is defined modulo $ 2\pi\/r $. A well defined operator is obtained by putting $V=e^{iX/r}$, which satisfies
\begin{align}
V\ket{n}=e^{\frac{i2\pi}{d}\/n}\ket{n} =q^{n}\ket{n},\hspace{20pt} \text{where} \hspace{20pt} q=e^{\frac{2\pi\/i}{d}} .
\end{align}
\\ \indent By the standard interpretation of quantum mechanics, V can be regarded as the unitary translation operator in momentum space. On the other hand, the translation operator in position space should be defined by 
\begin{gather}
U\ket{n}=\ket{n+1}, \, n=0,1,2,...,d-2 ,\\
U\ket{d-1}=\ket{0} .
\end{gather}
\\It follows that U and V satisfy \cite{schwinger1960unitary}
\begin{gather}
U^{d}=V^{d}=1, \\
VU=qUV, \hspace{20pt} \text{where} \hspace{20pt} q=e^{\frac{2\pi\/i}{d}} , \\
U^{\dagger}=U^{-1}, \;   V^{\dagger}=V^{-1},   
\end{gather}
\\which can be taken as the defining relations of quantum mechanics with d points in periodic position space and periodic momentum space. Eq. (6) is usually taken as the starting point of quantum mechanics in a space with a finite number of points. It can be shown that by taking the limit where the number of points is infinite, the Heisenberg commutation relation for $P$ and $X$ is obtained \cite{santhanam1976quantum}. Thus one can intuitively think that the correct choice of the Hamiltonian at the corresponding continuum limit will result in our calculations in this paper. However, until now, the formulation of the Hamiltonian has not been considered.
\\ \indent Taking eq. (6) as the starting point necessarily leads to periodicity in position space and momentum space. In this paper we will show that it is also possible to have quantum mechanics in a space with a finite number of points where the space is not periodic. 
\\ \indent In standard quantum mechanics both the translation operator in position space denoted by $U$ and the translation operator in momentum space denoted by $V$ are unitary. In terms of the momentum operator $P$ and the position operator $X$,
\begin{align}
U(a)&=e^{\frac{-iaP}{\hbar}} , \\
V(b)&=e^{\frac{ibX}{\hbar}} ,  \\
V(b)U(a)&=e^{\frac{iab}{\hbar}}U(a)V(b),
\end{align}
\\where $P$ and $X$ are well-defined hermitian operators. However for the discrete finite case only $U$ and $V$ are well-defined.
\\ \indent Bonatsos et al. \cite{bonatsos1994discretization} have considered the position and momentum operators for the q-deformed oscillator with $q$ being a root of unity. They have shown that the phase space of this oscillator has a lattice structure, which is a non-uniformly distributed grid. In contrast, in this paper we assume that the grid is uniform.
\\ \indent We define a deformed momentum operator 
\begin{gather}
\tilde{P}=\dfrac{2\pi\/\hbar}{ad}\frac{U^{-1/2}-U^{1/2}}{q^{1/2}-q^{-1/2}} , \hspace{20pt} \text{where}\hspace{20pt}
q=e^{\frac{2\pi\/i}{d}},\hspace{20pt} 
U=e^{\frac{-iaP}{\hbar}},
\end{gather}
\\where in the last equation $P$ is not well-defined since momentum is periodic. As $a\rightarrow\/0$ and $d\rightarrow\/\infty$ so that $ad=finite$
\begin{gather}
\tilde{P}=P+\mathcal{O}(P^{3}). 
\end{gather}
\\Although the half integer power in the exponent may look problematic, for the choice of the free particle Hamiltonian the half integer power is irrelevant. Thus our choice of the Hamiltonian is given by
\begin{align}
H=\dfrac{1}{2M} \tilde{P} ^{2}= \dfrac{\pi^{2}\hbar^{2}}{(ad)^{2}M} \dfrac{[2-(U+U^\dagger)]}{2sin^{2}(\pi/d)} ,
\end{align}
\\where $M$ is the mass of the particle.
\\ \indent For a Hamiltonian to be acceptable it should reduce to $\frac{p^{2}}{2m}$ in the continuum limit. Our choice of the Hamiltonian is not unique. However, it is the simplest Hamiltonian which gives $\frac{p^{2}}{2m}$ in the continuum limit because it contains the first power of $U$ and $U^\dagger$.
\section{Nonunitary translation operators}
\indent We would also like to address the quantum mechanics when the position space is not periodic.
We consider a position space of $d$ points where a particle located at position $ X=na $ is described by the ket vector $ \ket{n}, n=0,1,...,d-1 $ where
\begin{align}
X\ket{n}=na\ket{n} .
\end{align}
\\We regard the points $x=0$ and $x=(d-1)a$ as the end points of this discrete position space. We have to decide how the translation operator acts at the end points. For the right translation we define an operator $\ar$ whose action on the position eigenstates $\ket{n}$ is given by
\begin{gather}
\ar\ket{n}=\ket{n+1},\hspace{20pt} n=0,1,...,d-2 ,\\
\ar\ket{d-1}=0   .
\end{gather}
\\Similarly, the left translation operator will be denoted $\al$ which satisfies
\begin{gather}
\al\ket{n}=\ket{n-1},\hspace{20pt} n=1,...,d-1 ,\\
\al\ket{0}=0 .
\end{gather}
\\ \indent Note that this is the simplest kind of generalized oscillator which has a finite number of states. The most well-known of these generalized oscillators is the Biedenharn-Macfarlane oscillator \cite{biedenharn1989quantum,macfarlane1989q} with the spectrum 
\begin{align}
a^{\dagger}a=\dfrac{q^{N}-q^{-N}}{q-q^{-1}},
\end{align}
\\with real $q$. The spectrum also becomes well-defined when $q$ is a root of unity. The representations of the q-deformed oscillator algebra \cite{arik1976hilbert} with $q$ a root of unity are discussed in \cite{chung1997q}.
\\ \indent Taking the position eigenstates as orthonormal, the translation operators $\ar$ and $\al$, instead of being unitary, satisfy the algebra
\begin{gather}
\ar^{d}=\al^{d}=0 ,\\
\ar^{\dagger}=\al  ,\\
\agb  ,\\
\agi .
\end{gather}
\\ \indent Because our space has a finite number of points with $0\leq\/n\leq\/d-1$, 
\begin{align}
\ar^{d}\ket{n}=\al^{d}\ket{n}=0 .
\end{align} 
\\Thus eq.(20) is satisfied. One can easily define $\ar$ and $\al$ in terms of position eigenstates and can arrive at eq.(21) because these states are orthonormal.
\\ \indent Then we check the unitarity of the operators
\begin{gather}
\ar\ar^{\dagger}\ket{n}=\ar\al\ket{n}\neq\ket{n}, \hspace{20pt} \text{only when} \hspace{20pt} \ket{n}=\ket{0} ,\\
\al\al^{\dagger}\ket{n}=\al\ar\ket{n}\neq\ket{n}, \hspace{20pt} \text{only when} \hspace{20pt} \ket{n}=\ket{d-1} ,
\end{gather}
\\otherwise
\begin{gather}
\ar\ar^{\dagger}\ket{n}=\ar\al\ket{n}=\ket{n},\hspace{20pt} n=1,2,...d-1,   \\
\al\al^{\dagger}\ket{n}=\al\ar\ket{n}=\ket{n}, \hspace{20pt} n=0,1,...d-2 .
\end{gather}
\\This means that we should have an additional term that breaks unitarity. This is the second term on the right hand side of eqs.(22) and (23). One can easily show that eqs. (22) and (23) are not only satisfied at the end states but also satisfied at intermediate states. The minimal set of relations which define the algebra generated by $\ala$ and $\ara$ is given by 
\begin{align}
\ala\ara=1-\Pm{d-1}{d-1}, \hspace{20pt} \text{and} \hspace{20pt} \ala^{d}=0,  \nonumber
\end{align}
\\where $\ar=\ara$ and $\al=\ala$.
\\ \indent Now we will show that these relations imply a unique representation with the desired properties and lead to a set of relations satisfied by $\ala$ and $\ara$. We define two useful operators
\begin{gather}
P_{0} = 1 ,\\
P_{n} = \Pmn .
\end{gather}
\\If we can show that
\begin{align}
P_{n}P_{m}=P_{m} , \hspace{25pt}  m\geq\/n,
\end{align}
\ we will have the inclusion relation 
\begin{align}
P_{n}^{2}=P_{n} .
\end{align}
Thus we can consider them as the projection operators. We will prove eq.(31) by induction. For $ n=1 $, we write
\begin{align}
P_{1}P_{m}&=\Pmb\Pmm   \nonumber \\
&=\ara(\ala\ara)\Pm{m-1}{m} 
\end{align}
\\using eq. (22)
\begin{align}
P_{1}P_{m}=\ara(1-P_{d-1})\Pm{m-1}{m} ,
\end{align}
\\and using eq. (20) 
\begin{align}
P_{1}P_{m}=\Pmm=P_{m}.
\end{align}
\\For $ n=l+1 $, we have
\begin{align}
P_{l+1}P_{m}&=\ara\/P_{l}(\ala\ara)P_{m-1}\ala    \nonumber \\
&=\ara\/P_{l}(1-P_{d-1})P_{m-1}\ala   \nonumber \\
&=\ara\/P_{l}P_{m-1}\ala-\ara\/P_{l}P_{d-1}P_{m-1}\ala ,\hspace{25pt} \text{where} \hspace{25pt} l+1\leq\/m .
\end{align}
\\Assuming eq.(31) holds for $n=l$ and remembering $l+1\leq\/m$ implies $l\leq\/m-1$
\begin{align}
P_{l+1}P_{m}=\ara\/P_{m-1}\ala-\ara\/P_{d-1}P_{m-1}\ala ,
\end{align}
\\using $\ara\/P_{d-1}=0$
\begin{align}
P_{l+1}P_{m}=\ara\/P_{m-1}\ala=P_{m} .
\end{align}
\\Thus the proof is complete.
\\ \indent Before going further, we will supply two more useful equations
\begin{align}
P_{m}\ara&=\Pm{m}{m-1}(\ala\ara)  \nonumber \\
&=\ara\/P_{m-1}(1-P_{d-1}) \nonumber \\
&=\ara\/P_{m-1}-\ara\/P_{d-1} \nonumber \\
&=\ara\/P_{m-1} ,
\end{align}
\\and
\begin{align}
\ala\/P_{m}&=(\ala\ara)P_{m-1}\ala  \nonumber  \\
&=(1-P_{d-1})P_{m-1}\ala  \nonumber \\
&=P_{m-1}\ala-P_{d-1}\ala \nonumber \\
&=P_{m-1}\ala .
\end{align}
\\By replacing $m$ with $m+1$, we obtain
\begin{align}
P_{m}\ala=\ala\/P_{m+1} .
\end{align}
\\ \indent Now we will prove
\begin{align}
\Pmt{n}{n}=1-P_{d-n} ,
\end{align}
\\which implies the last part of our algebra eq.(23) when $n=d-1$
\begin{align}
\Pmt{d-1}{d-1}=1-\ara\ala  . \nonumber
\end{align}
\\Our method is proof by induction. For $n=1$
\begin{align}
\ala\ara=1-P_{d-1} ,
\end{align}
\\which is eq.(22).
\\For $n=k+1$, we have
\begin{align}
\Pmt{k+1}{k+1}=\ala(\Pmt{k}{k})\ara. 
\end{align}
\\For $n=k$ we assume $\Pmt{k}{k}=1-P_{d-k}$. By substuting this in the last equation, we obtain
\begin{align}
\Pmt{k+1}{k+1} &= \ala(1-P_{d-k})\ara \nonumber \\
&=\ala\ara-\ala\/P_{d-k}\ara .
\end{align}
\\Using eq. (22) for the first term on the RHS and eq. (41) for the second term on the RHS, we get
\begin{align}
\Pmt{k+1}{k+1}=1-P_{d-1}-P_{d-k-1}\ala\ara ,
\end{align}
\\using eq.(22) once more for the term $\ala\ara$,
\begin{align}
\Pmt{k+1}{k+1}=1-P_{d-1}-P_{d-k-1}(1-P_{d-1}) ,
\end{align}
\\remembering $P_{n}P_{m}=P_{m}$ for $m\geq\/n$, which was our first proved equation
\begin{align}
\Pmt{k+1}{k+1}&=1-P_{d-1}-P_{d-k-1}+P_{d-1} \nonumber \\
\Pmt{k+1}{k+1}&=1-P_{d-(k+1)}  \hspace{10pt} \text{Q.E.D.} .
\end{align}
\section{Eigenvalues and eigenfunctions of the Hamiltonian for the nonperiodic space} 
\indent In a manner similar to eq. (13), we choose the Hamiltonian for the non-periodic case
\begin{align}
H= \dfrac{\pi^{2}\hbar^{2}}{(ad)^{2}M} \dfrac{[2-(\ala+\ara)]}{2sin^{2}(\pi/d)} .
\end{align}
\\We now calculate the eigenvalues and eigenvectors of $(\ala+\ara)$
\begin{gather}
\ket{\lambda}=\sum\limits_{n=0}^{d-1}\lambda_{n}\ket{n} , \\
(\ala+\ara)\ket{\lambda}=\sum\limits_{n=0}^{d-1}(\lambda_{n}\ket{n-1}+\lambda_{n}\ket{n+1})=\sum\limits_{n=0}^{d-1}\lambda\lambda_{n}\ket{n} ,
\end{gather}
\\so
\begin{align}
\lambda_{n+1}+\lambda_{n-1}=\lambda\lambda_{n} , \hspace{15pt} \text{with} \hspace{15pt} \lambda_{-1}=0 , \hspace{15pt} \text{and} \hspace{15pt} \lambda_{d}=0 .
\end{align}
\\We obtain the general solution $\lambda_{n}=Ae^{in\theta}+Be^{-in\theta}$, where $cos\theta=\lambda/2$. Using the boundary conditions $\lambda_{-1}=0$ and $\lambda_{d}=0$, we get $B=-Ae^{-2i\theta}$ and $\theta=\dfrac{\pi\\m}{d+1}$ where $m=1,2,...,d$. $m=0$ is excluded because it does not yield a non-zero eigenvector. Thus for each value of $n$, we obtain $m$ different eigenvalues, given by
\begin{align}
\lambda_{n}^{(m)}&=A_{m}(e^{i\alpha nm}-e^{-2i\alpha m}e^{-i\alpha nm})  \hspace{15pt} \text{where} \hspace{15pt} \alpha=\pi/(d+1) ,\\
&=A_{m}2ie^{\frac{-im\pi}{1+d}}sin[\dfrac{m(1+n)\pi}{1+d}] .
\end{align}
\\If we normalize $\ket{\lambda^{(m)}}$, we obtain
\begin{align}
A_{m}=\sqrt{\dfrac{1}{2(d+1)}} .
\end{align}
\\As a result the eigenvectors are written as
\begin{align}
\ket{\lambda^{(m)}}=\dfrac{2ie^{\frac{-im\pi}{1+d}}}{\sqrt{2(d+1)}}\sum\limits_{n=0}^{d-1}sin[\dfrac{m(1+n)\pi}{1+d}]\ket{n} ,
\end{align}
\\and the corresponding eigenvalues are given by
\begin{align}
\lambda^{(m)}=2cos\theta=2cos(\frac{\pi\/m}{d+1}) \hspace{15pt} \text{with} \hspace{15pt} m=1,2,...d .
\end{align}
\\ \indent Now, it is time to compare what we have found with the usual quantum particle in an infinite one-dimensional square well. We calculate the eigenvalues of the Hamiltonian, by applying eq.(49) to eigenvectors given by eq.(56). We find
\begin{align}
E_{m}=\dfrac{2\pi^{2}\hbar^{2}}{(ad)^{2}M}{\dfrac{sin^2[\frac{\pi\/m}{2(d+1)}]}{sin^2(\frac{\pi}{d})}} \hspace{15pt} \text{where} \hspace{15pt} m=1,2,...,d.
\end{align} 
\\In the continuum limit $a\rightarrow\/0$, $d\rightarrow\infty$ and $a(d-1)=L$, these energy eigenvalues become
\begin{align}
 E_{m}=\dfrac{\pi^2\/\hbar^2\/m^2}{2ML^2} , 
\end{align}
\\which are the same energy eigenvalues in the case of a particle confined in an infinite one dimensional square well of width L.
 The wave functions are given by 
\begin{align}
\psi_{m}(n)=C^{(m)}_{n} \braket{n}{\lambda^{(m)}} = C^{(m)}_{n}\sqrt{\dfrac{2}{d+1}}sin[\dfrac{m(1+n)\pi}{d+1}] .
\end{align}
\\We can easily find $C^{(m)}_{n}$ by normalization of the wave function,
\begin{align}
\sum\limits_{n=0}^{d-1}\abs{\psi_{m}(n)}^{2}a=1, \hspace{20pt} C^{(m)}_{n}=\dfrac{1}{\sqrt{a}} .
\end{align}
\\Thus the wave functions can be written as
\begin{align}
\psi_{m}(n)=\sqrt{\dfrac{2}{(d+1)a}}sin[\dfrac{m(1+n)\pi}{d+1}] \hspace{15pt} \text{where} \hspace{15pt} m=1,2,...,d.
\end{align}
\\In the continuum limit the wave function becomes
\begin{align}
 \psi_{m}(x)=\sqrt{\dfrac{2}{L}}sin[\dfrac{m\pi\/x}{L}] ,
\end{align}
\\which agrees with the wave functions of a square well in the interval $0\leq\/x\leq\/L$ .
\section{Eigenvalues and eigenfunctions of the Hamiltonian for the periodic space}
\indent For the periodic case, we need to find eigenvalues of $(U+U^{\dagger})$ using the position basis. Our choice is the same as eq. (50). If we apply $(U+U^{\dagger})$ to this eigenvector, we will have
\begin{align}
\lambda_{n+1}+\lambda_{n-1}=\lambda\lambda_{n}  \hspace{15pt} \text{with} \hspace{15pt} \lambda_{0}=\lambda_{d} .
\end{align} 
\\Since the equation is linear, we can superpose linearly independent solutions to find the general solution. Since we desire the linearly independent solutions to be real in order to be able to plot the wave function, we choose $ \lambda_{n} $ as
\begin{align}
\lambda_{n}=Ccos(n\theta)+Dsin(n\theta) ,
\end{align}
\\which satisfies eq.(64) with $\lambda=2cos\theta$. If we impose the boundary condition $\lambda_{n}=\lambda_{d+n}$, we obtain $\theta=\dfrac{2\pi\/m}{d}$ with $m=0,1,...,d-1$ and 
\begin{align}
\ket{\lambda^{(m)}}&=\sum\limits_{n=0}^{d-1}{[Ccos(\frac{2\pi\/nm}{d})+Dsin(\frac{2\pi\/nm}{d})]}\ket{n} . 
\end{align}
\\ At this point there are options regarding the choice of $C$ and $D$. By introducing a phase angle we can write this equation as
\begin{align}
\ket{\lambda^{(m)}}&=\sum\limits_{n=0}^{d-1}C_{m}cos(\frac{2\pi\/nm}{d}-\gamma_{m})\ket{n} \hspace{20pt} \text{where}\hspace{20pt}
\gamma_{m}=\frac{2\pi\/n_{0}m}{d} .
\end{align}
By selecting $n_{0}$ and $n=-\frac{d-1}{2},-\frac{d-1}{2}+1,...,\frac{d-1}{2}$, the solutions become parity eigenstates. Thus parity corresponds to $n\rightarrow -n$. As a result, $n_{0}=\frac{d-1}{2}$ and for normalized eigenvectors we get
\begin{align}
C_{m} =
\begin{cases}
\sqrt{\dfrac{1}{d}} & \quad \text{if} \hspace{10pt} m=0,\\
\sqrt{\dfrac{2}{d}} &  \quad \text{if} \hspace{10pt} m\neq\/0.
\end{cases}
\end{align}
\\The final form of eigenvectors is given by
\begin{align}
\ket{\lambda^{(m)}}=C_{m}\sum\limits_{n=0}^{d-1}cos[\dfrac{2\pi\/nm}{d}-\dfrac{2\pi(d-1)m}{2d}]\ket{n} ,
\end{align}
\\and the eigenvalues are given by
\begin{align}
\lambda^{(m)}=2cos\theta=2cos(\frac{2\pi\/m}{d}) \hspace{10pt} \text{where} \hspace{10pt}
m=
\begin{cases}
0,1,...,int(\dfrac{d-1}{2}) & \quad \text{for positive parity states} ,\\
1,2,...,int(\dfrac{d}{2}) &  \quad \text{for negative parity states}.
\end{cases}
\end{align} 
\\Hence we find the energy eigenvalues as 
\begin{align}
E_{m}=\dfrac{2\pi^{2}\hbar^{2}}{(ad)^{2}M}{\dfrac{sin^2[\frac{m\pi}{d}]}{sin^2(\frac{\pi}{d})}} .
\end{align} 
\\In the continuum limit, we set $ad=L$ and find
\begin{align}
E_{m}=\dfrac{2\pi^2\/\hbar^2\/m^2}{ML^2},
\end{align}
\\which are the same as energy eigenvalues in the case of particle in a box with periodic boundary conditions.
\\ \indent We continue by obtaining the wave functions. Applying the same method that we used for the nonperiodic case, we find that the ground state is unique and given by
\begin{align}
\psi_{0}(n)=\sqrt{\dfrac{1}{da}}.
\end{align}
\\For odd $d$ the excited states are doubly degenerate, whereas for even $d$ the excited states are doubly degenerate except the highest energy state. The eigenvalues of the position operator are given by
\begin{align}
X\ket{n}=an\ket{n}, \hspace{20pt} n=-\dfrac{d-1}{2},-\dfrac{d-1}{2}+1,...,\dfrac{d-1}{2}.
\end{align}
Then the positive parity excited states are given by the wave functions
\begin{align}
\psi^{+}_{m}(n)=\sqrt{\dfrac{2}{da}}cos(\dfrac{2\pi\/mn}{d}), \hspace{20pt} m=1,2,...,int(\dfrac{d-1}{2}),
\end{align}
\\and the negative parity excited states are given by the wave functions
\begin{align}
\psi^{-}_{m}(n)=\sqrt{\dfrac{2}{da}}sin(\dfrac{2\pi\/mn}{d}), \hspace{20pt} m=1,2,...,int(\dfrac{d}{2}) . 
\end{align}
\\ \indent If one plots the wave functions, it is observed that the wave functions of the omitted $m$ values are the same as the wave functions of the indicated $m$ values. The total number of states, including the ground state, is $d$. 
\\ \indent At the continuum limit, the ground state wave function, the even parity wave functions and the odd parity wave functions respectively become
\begin{align}
\psi^{+}_{0}(x)&=\sqrt{\dfrac{2}{L}}, \\
\psi^{+}_{m}(x)&=\sqrt{\dfrac{2}{L}}cos[\dfrac{2m\pi\/x}{L}] , \\
\psi^{-}_{m}(x)&=\sqrt{\dfrac{2}{L}}sin[\dfrac{2m\pi\/x}{L}] ,
\end{align}
\\which are the even parity solutions and the odd parity solutions for a particle in a box in an interval $-L/2\leq\/x\leq\/L/2$ with periodic boundary conditions.
\section{Discussion}
\indent The $d=2$ case gives the fermionic oscillator. For this case $\ara$ and $\ala$ correspond to the creation and annihilation operators for a single fermionic degree of freedoom. For the periodic case Schwinger \cite{schwinger2001quantum} has shown that for $d=2$, $U$ and $V$ in fact generate the Pauli algebra. Our work shows that even for the non-periodic case, $d=2$ leads to Pauli-matrices through the relations,
\begin{align}
\sigma_{1}&=\ara+\ala, \\
\sigma_{2}&=i(\ara-\ala), \\
\sigma_{3}&=\ala\/\ara-\ara\/\ala.
\end{align}
\\ \indent We should also note that a universe with a finite number of points may be a physical reality. The size of the visible universe is $ 10^{27}$ m, whereas the smallest classical length is the Planck length, $10^{-35}$ m. This means that the physical space continuum can be regarded as consisting of $d_{universe}/l_{planck}=10^{62}$ points lying along one dimension. Thus, it may be that what we call the space continuum can be described by quantum mechanics with $d=10^{62}$ points along one dimension. Another possibility is that in a Klauza-Klein like theory \cite{viet1996noncommutative} the internal space can consist of a finite number of points.

\section*{Acknowledgements}
\indent We would like to acknowledge fruitful discussion with Tonguc Rador and Teoman Turgut.

\bibliographystyle{unsrt}
\bibliography{bibfile}

\begin{thebibliography}{10}

\bibitem{john1955}
John~Von Neumann.
\newblock {\em Mathematical foundations of quantum mechanics}.
\newblock Number~2. Princeton university press, 1955.

\bibitem{0034-4885-67-3-R03}
A~Vourdas.
\newblock Quantum systems with finite hilbert space.
\newblock {\em Reports on Progress in Physics}, 67(3):267, 2004.

\bibitem{schwinger1960unitary}
Julian Schwinger.
\newblock Unitary operator bases.
\newblock {\em Proceedings of the national academy of sciences of the United
  States Of America}, 46(4):570, 1960.

\bibitem{santhanam1976quantum}
TS~Santhanam and AR~Tekumalla.
\newblock Quantum mechanics in finite dimensions.
\newblock {\em Foundations of Physics}, 6(5):583--587, 1976.

\bibitem{bonatsos1994discretization}
Dennis Bonatsos, C~Daskaloyannis, Demosthenes Ellinas, and Amand Faessler.
\newblock Discretization of the phase space for a q-deformed harmonic
  oscillator with q a root of unity.
\newblock {\em Physics Letters B}, 331(1):150--156, 1994.

\bibitem{biedenharn1989quantum}
LC~Biedenharn.
\newblock The quantum group suq (2) and a q-analogue of the boson operators.
\newblock {\em Journal of Physics A: Mathematical and General}, 22(18):L873,
  1989.

\bibitem{macfarlane1989q}
AJ~Macfarlane.
\newblock On q-analogues of the quantum harmonic oscillator and the quantum
  group su (2) q.
\newblock {\em Journal of Physics A: Mathematical and general}, 22(21):4581,
  1989.

\bibitem{arik1976hilbert}
M~Arik and DD~Coon.
\newblock Hilbert spaces of analytic functions and generalized coherent states.
\newblock {\em Journal of Mathematical Physics}, 17(4):524--527, 1976.

\bibitem{chung1997q}
WS~Chung.
\newblock q-oscillators with q a root of unity.
\newblock 1997.

\bibitem{schwinger2001quantum}
Julian Schwinger and Berthold-Georg Englert.
\newblock Quantum mechanics: Symbolism of atomic measurements.
\newblock page~73, 2001.

\bibitem{viet1996noncommutative}
Nguyen~Ai Viet and Kameshwar~C Wali.
\newblock Noncommutative geometry and a discretized version of kaluza-klein
  theory with a finite field content.
\newblock {\em International Journal of Modern Physics A}, 11(03):533--551,
  1996.

\end{thebibliography}
\end{document}